\documentclass[11pt]{article}
\usepackage{moriond,epsfig}

\def\be{\begin{equation}}
\def\ee{\end{equation}}
\def\bea{\begin{eqnarray}}
\def\eea{\end{eqnarray}}

\begin{document}
\vspace*{4cm}
\title{DETERMINATION OF NEUTRINO MASSES, PRESENT AND FUTURE}

\author{ J.-L.~Vuilleumier }

\address{Institut de physique, A.-L. Breguet 1, 
                   CH-2000 Neuch\^atel, Switzerland}

\maketitle\abstracts{Oscillation experiments show that neutrinos have
masses. They however only determine the neutrino mass
differences. Information on the absolute masses can be obtained by
studying the kinematics in weak decays, or by searching for
neutrinoless double beta decay. Recent results are reviewed, as well
as future projects.}

\section{Introduction}
The lagrangian for the neutrinos may contain Dirac
mass terms $m^{D}$, as for all other particles. In addition, Majorana
mass terms $m^{L}$ and $m^{R}$ for left respectively right handed
neutrinos, can appear: 
\[
{\cal{L}} =\sum_{\ell,\ell'} ( m_{\ell \ell'}^{D}\overline{\nu}_{\ell
    L}\nu_{\ell'R} + m_{\ell \ell'}^{L}\overline{\nu}_{ \ell
    R}^{c}\nu_{\ell' L} + m_{\ell \ell'}^{R}\overline{\nu}_{
    \ell L}^{c}\nu_{\ell' R} )  +h.c. 
\]
The Dirac mass terms conserve the total lepton number ($\Delta L=0$),
while the Majorana terms break it by two units ($\Delta L=2$). With
them lepton number violating processes such as neutrinoless double
beta ($\beta\beta0\nu$) decay may occur. In any case the mass matrices
are not in the most general case diagonal in the flavor $\ell$
($\ell=e,\;\mu\;\tau$...).  The flavor eigenstates $\nu_{\ell}$ are
then, assuming $N$ generations of neutrinos, superpositions of $2N$
Majorana mass eigenstates $\nu_{i}$ with mass $m_{i}$.
With Dirac mass terms only this reduces to a sum extending over $N$ Dirac 
mass eigenstates $\nu_{i}$:
\[
\nu_{\ell} = \sum_{i=1}^{N} U_{\ell i} \nu_{i}. \label{eq:massmix}
\]
and the flavor eigenstates $\nu_{\ell}$ are Dirac as well. This
equation also holds in the case of Majorana mass terms only. All
eigenstates are then Majorana, with $\nu_{\ell}$ representing the left
handed Majorana flavor eigenstates.

With an off-diagonal mass matrix, neutrinos will undergo oscillations
and transitions. With Dirac mass terms only, or Majorana mass terms
only, because of the last equation, the oscillations or transitions
take place between flavor eigenstates only.

Oscillation experiments are sensitive to the 
mixings $ U_{\ell i}$ and the squared mass differences
\[
\Delta m_{ik}^{2} = m_{i}^{2} - m_{k}^{2}.
\]
Experiments with solar neutrinos determine a value $\Delta
m^2_{sol}\sim 5\cdot10^{-5}$~eV$^2$ and a large, but not maximum,
mixing \cite{SNO03}. Atmospheric neutrino experiments yield a second
mass squared difference $\Delta m^2_{atm}\sim 2.5\cdot10^{-3}$~eV$^2$
and a mixing consistent with the maximal value \cite{SK03}. These experiments
cannot however determine the absolute masses.

This article is devoted to two other types of experiments providing 
informations on the absolute masses. First is the study of the kinematics 
in weak decays, such as
$^3$H$\rightarrow ^3$He$ + e^- + \overline{ \nu}_e$. 
These measure directly, in principle, the mixings
$U_{\ell, i}$ and the masses $m_i$. And then additional information can 
be gained by studying $\beta\beta0\nu$ decay, which measures an effective 
mass, exactly zero with Dirac masses alone, and taking the shape:
\[
|\langle m_{\nu}\rangle | = |\sum_{i=1}^{N}U_{ei}U_{ei}m_i| \label{eq:meff}
\]
with Majorana masses alone. 
\begin{figure*}[htb]
\begin{center}
\hspace*{-0.cm}
\epsfig{file=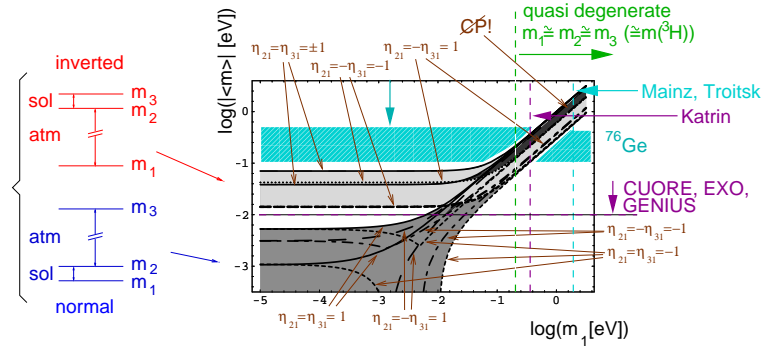,width=10.cm}
\caption{Region allowed by oscillation experiments in the $\langle m_{\nu}
\rangle$ (effective double beta decay mass) vs. $m_1$ (lightest
neutrino mass). Existing limits and
projected sensitivities are indicated.}
\label{fi:DBDg}
\end{center}
\end{figure*}

Moreover it must be mentioned that the study of cosmic microwave
background (CMB) anisotropies  and large scale structure (LSS) surveys
of galaxies constrain the sum of all light neutrino masses.  Relic
Big-Bang neutrinos with too large masses would wash out these
structures. A careful study of the CMB anisotropies measured by the
WMAP satellite and of the 2dFGRS survey, assuming three
neutrino families ($N=3$), yields the limit $\sum_{i=1}^3 m_i<2.12$~eV
at 95 \% CL \cite{Hann03}.  It can be brought down to 1.01 eV by
including additional input from cosmology.
   
The solar and atmospheric data are described well with
three neutrino families. In that case, and assuming pure Majorana masses, 
the effective mass reduces to
\[
|\langle m_{\nu}\rangle |  = |m_1|U_{e1}|^2 + m_2|U_{e2}|^2e^{i\alpha_{21}} 
+   m_3|U_{e3}|^2e^{i\alpha_{31}}|,
\]
where $\alpha_{21}$ and $\alpha_{31}$ are CP violating phases 
\cite{BilP01,PasP02}. In case of CP 
conservation they are such that
\[
\eta_{21}=e^{i\alpha_{21}}=\pm1,\;\;\eta_{31}=e^{i\alpha_{31}}=\pm1.
\]

The aforementioned solar and atmospheric neutrino data, with their
uncertainties, constrain two of the mixings, and the two squared mass
differences. The third mixing is constrained by the Chooz reactor
experiment \cite{Choo03}. All this, for a given mass scale, usually
taken as the mass $m_1$ of the lightest neutrino, constrains in turn
the allowed range of $|\langle m_{\nu}\rangle |$
\cite{BilP01,PasP02,EllV02}.  This is illustrated in fig.
\ref{fi:DBDg}, taken from ref. \cite{PasP02}, where the allowed
regions in the $|\langle m_{\nu}\rangle |$ vs. $m_1$ plane are
shown. For $m_1$ above 0.1 eV, the masse differences are small
compared to the absolute masses. It is the quasi degenerate
region. All masses take nearly the same value
($m_1\simeq m_2 \simeq m_3$), which is the quantity measured in kinematic
experiments. Because of possible cancellations, depending on the
CP-phases, $|\langle m_{\nu}\rangle |$ can be smaller than $m_1$, by
about a factor 5. For $m_1$ below 0.01 eV the mass differences become large
compared to $m_1$ itself. Here one
distinguishes two scenarios. In the first one, corresponding to the
normal hierarchy, solar neutrinos oscillate between the two light
neutrinos, and atmospheric neutrinos between a light and a heavy
one. In that case $|\langle m_{\nu}\rangle |$ is always less than
$10^{-2}$~eV. In the second scenario, corresponding to inverse
hierarchy, the solar neutrinos oscillate between the heavy neutrinos,
and atmospheric neutrinos between a heavy one and the light one. With
this, $|\langle m_{\nu}\rangle |$ is always larger than
$10^{-2}$~eV. 

The study of double beta decay and of the kinematics in weak decays
thus gives the possibility to discriminate between these scenarios,
and to fix the absolute mass scale. In particular, if $m_1$ is less
than 0.01 eV, double beta decay can find out about the hierarchy,
normal or inverted.  For this it is however necessary to push the
sensitivity down to the 0.01 eV level. Even if one can hope that
ongoing and future oscillation experiments, in particular Kamland
\cite{Kaml03}, will further constrain the allowed areas in
fig. \ref{fi:DBDg}.

In the following we are going to discuss existing bounds on neutrino
masses, and look how far the sensitivity could extend in next
generation experiments.
\section{Kinematics in weak decays}
\subsection{Electrostatic spectrometers}
\begin{figure}[htb]
\begin{center}
\hspace*{-0.cm}
\epsfig{file=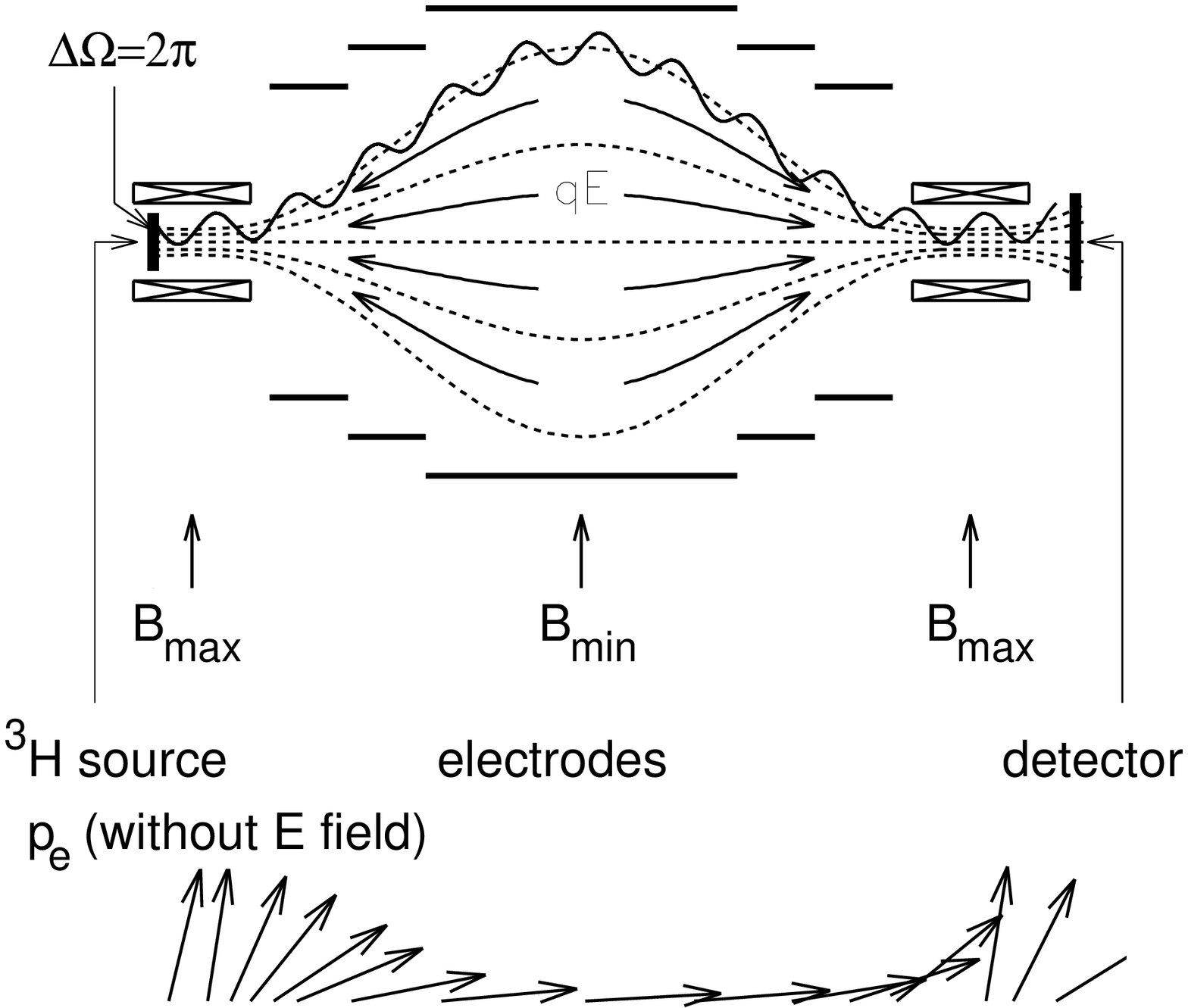,width=7.cm}
\epsfig{file=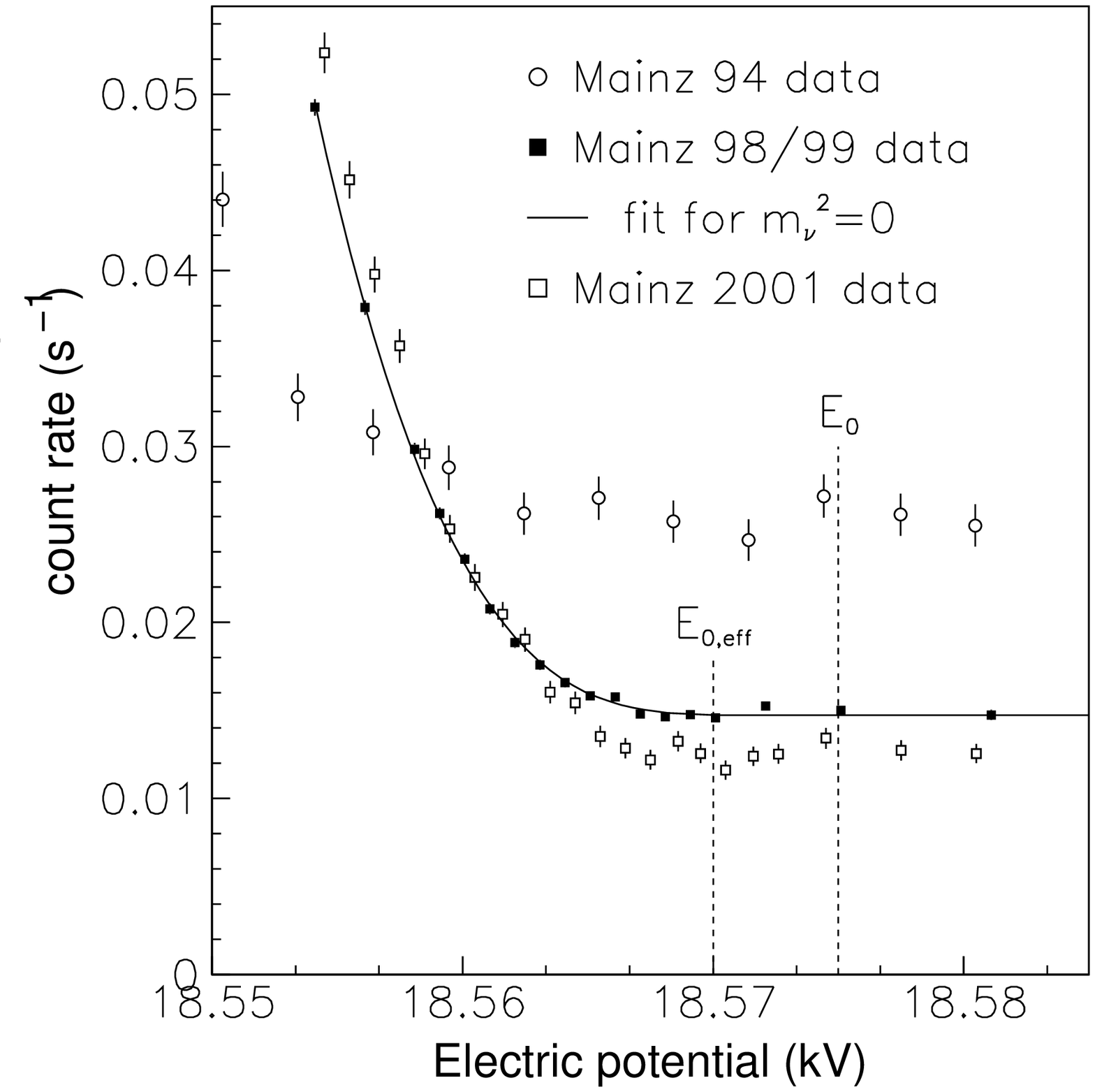,width=7.cm}
\caption{Left: principles of operation of an integral electrostatic
spectrometer; evolution of the momentum of an electron spiraling
in the magnetic field for no electric field; right: the Mainz 
$^3$H spectra near end-point.}
\label{fi:inspe}
\end{center}
\end{figure}

For many years the best limits on the mass of the neutrino admixed most
to the electron neutrino have come from the
study of tritium decay $^3$H$ \rightarrow ^3$He$ + e^- + \overline{ \nu}_e$
with an end-point energy of $E_0$=18.6 keV. One looks for a distortion of
the beta spectrum near the end-point due to the masses, 
as described in more details in ref. \cite{Wein03}.  Here the
important parameters are high energy resolution, large acceptance
since only a tiny fraction of the emitted electrons fall in the region
of interest, and low background. The source must be such as to
minimize energy losses. Presently the best results come from integral
electrostatic spectrometers, built and operated in Troitsk
\cite{Lob99} and Mainz \cite{Main02}. These instruments consist in two
superconducting coils, generating a magnetic field, as shown in
fig. \ref{fi:inspe}. The source is placed at the entrance of one of
the coils. The electrons emitted in the forward half-sphere spiral
toward the detector, located behind the second coil. As they move to
regions with a weaker magnetic field, the transverse momentum changes
into longitudinal motion. Electrodes are used to create an electric
potential barrier $U$. Only electrons with an energy above $eU$ can
pass the barrier, and are accelerated toward the detector. Scans are
performed varying the potential $U$, providing the integral electron
spectrum. The energy resolution is given by
\[
\frac{\Delta E}{E} = \frac{B_{min}}{B_{max}}.
\]

The Troitsk device uses a gaseous $^3$H$_2$ source, which has the
advantage of minimal energy losses. The energy resolution is 3.5
eV. Early spectra showed an anomaly near end points, fits with one
single mass yielding a negative squared mass. This is now
believed to be due to an experimental artifact, and was eliminated in
the most recent spectra. Older data are necessary however to minimize
the statistical errors. The negative squared mass can be eliminated by
adding in the fit a step to the integral spectrum, ending at an energy
below $E_0$, which was found to vary in time. With this a negative squared mass
consistent with zero is obtained, leading to the limit $ m_{\nu}< 2.2$~eV at
95 \% CL.

The Mainz system has an energy resolution of 4.8 eV. It uses a frozen
$^3$H$_2$ source. The experiment went through several upgrades, which led
in particular to a lower background, as shown in
fig. \ref{fi:inspe}. Distortions leading here also to negative squared masses
were understood as being due to a roughening of the source film and
were eliminated by substantially reducing the operating temperature. The last
years of operation give a negative squared mass §consistent with zero,
and to the same limit
\[
m_{\nu}< 2.2\mbox{  eV at 95 \% CL,}
\]
which can thus be considered presently the best upper bound on the neutrino
mass, ruling out part of the quasi degenerate region in fig. \ref{fi:DBDg}.

The Mainz and the Troitsk groups consider that they have
exhausted the potential of their instruments. They have joined efforts
and, along with a few additional teams, have undertaken the
construction of Katrin, a much larger device \cite{Wein03}. The energy
resolution should be of order 1 eV. A small spectrometer will be placed 
in front of the main one, filtering out all the electrons clearly
below the end-point, reducing the background considerably. A
gaseous and a frozen $^3$H source will be used, helping in pinning down 
the systematics associated
with either source. With all this it is hoped that the sensitivity to the
neutrino mass will extend down to 0.35 eV.
\subsection{Bolometers}
Cryogenic bolometers, proposed first by the Genova group,
may turn out to be fairly competitive in searching for the neutrino mass.

In these, the source, in crystalline form, is the detector medium
itself. The entire energy deposited by an event ends up in heat. The
crystal is operated at low temperature, so that the corresponding
temperature increase is large, and can be measured with a thermistor
glued to the crystal.  In principle, no corrections need to be made
for final state or source effects.

\begin{figure}[htb]
\begin{center}
\hspace*{-0.cm}
\epsfig{file=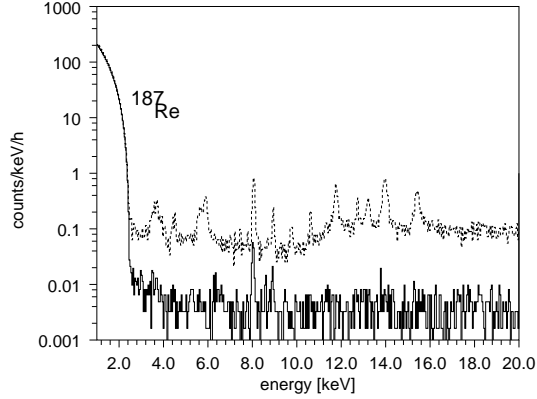,width=7.cm}
\caption{The $^{187}$Re spectrum measured with the Milano bolometers, with and without calibration source.}
\label{fi:MiRe}
\end{center}
\end{figure}

The Genova group \cite{Galea98} has demonstrated the feasibility of
using metallic rhenium crystals to look for the decay
$^{187}$Re$\rightarrow^{187}$Os$+e^-+\overline{\nu}_e$ (natural
abundance of $^{187}$Re 62 \%). One advantage is the low end point
energy ($E_0\simeq$2.5~keV), which makes that a larger fraction of
events falls in the region of interest. But the most recent results
have been produced by the Milano group \cite{ArnB03}. An array of 10 AgReO$_4$
crystals with a mass of 250-300~$\mu$g each has been built and operated for
a longer period of time. The measured spectra with and without
calibration source are shown in fig. \ref{fi:MiRe}.  The energy
resolution is 28 eV on average at 2.5 keV. The measured spectrum was
found to be in good agreement with a vanishing neutrino mass, and the
limit $m_{\nu}<21.7$ eV at 90 \% confidence was derived. This is limited
by statistics, and by the energy resolution. But clearly the method
still has a lot of potential.
\section{Double beta decay}
We are going to discuss neutrinoless and two neutrino nuclear double
beta decay. The two neutrino mode $\beta\beta2\nu$
$(A,Z) \longrightarrow (A,Z+2) + e^- + e^- +\overline{\nu}_e +
\overline{\nu}_e$ conserves the lepton number and is allowed in the
standard model. The half life is given by
\[
\left(T^{2\nu}_{1/2}\right) ^{-1}=G^{2\nu}(E_0,Z)|M^{2\nu}_{GT}|^2
\]
with $G^{2\nu}(E_0,Z)$ an exactly calculable phase space factor
depending only on the nuclear charge $Z$ and the total energy released
in the decay $E_0$, and $M^{2\nu}_{GT}$ a matrix element.

Neutrinoless double beta decay $\beta\beta0\nu$
$
(A,Z) \longrightarrow (A,Z+2) + e^- + e^-
$
breaks the lepton number by two units and requires Majorana
masses. The half life depends directly on the effective mass $\langle
m_{\nu}\rangle$:
\[
\left(T^{0\nu}_{1/2}\right)
^{-1} = 
  G^{0\nu}(E_0,Z)\left|M^{0\nu}_{GT}-\frac{g_V^2}{g_A^2}
M^{0\nu}_F\right|^2 |\langle m_{\nu}\rangle|^2, 
\]
with again $G^{0\nu}(E_0,Z)$ a phase space factor, $g_V$ and $g_A$ 
the vector and axial vector coupling constants, and $M^{0\nu}_{GT}$
and $M^{0\nu}_F$ matrix elements which are not the same as those of
$\beta\beta2\nu$, but which depend on the same nuclear physics.

\begin{figure}[htb]
\begin{center}
\hspace*{-0.cm}
\epsfig{file=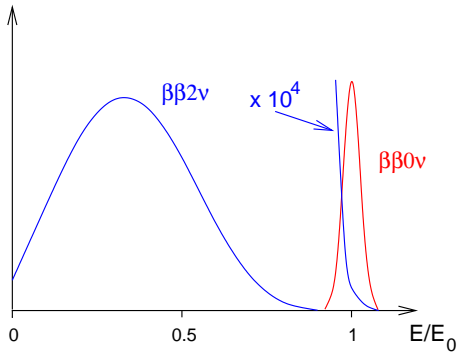,width=6.cm}
\caption{The expected energy distributions of the sum energy of the
two neutrinos for $\beta\beta 0\nu$ and $\beta\beta 2\nu$ decay, smeared
by energy resolution (gaussian with $\sigma (E)/E$=2$\%$).
\label{fi:bb2n}}
\end{center}
\end{figure}
Experimentally the two modes can be distinguished by measuring the
total energy carried away by the two electrons. In ($\beta\beta2\nu$)
it follows a smooth distribution peaking at about 1/3 $E_0$, while in
($\beta\beta0\nu$) it is equal to $E_0$. In either case the
theoretical distributions are smeared by the instrumental resolution,
as shown in figure \ref{fi:bb2n}. Many isotope candidates have been
explored over the past years. The criterias of selection are
high energy release $E_0$, corresponding to a large phase space,
and placing the search for $\beta\beta0\nu$ decay in a region with
lower background, and large natural abundance, which facilitates
enrichment.
\subsection{Present experiments}
$\beta\beta2\nu$ decay has now been seen in many nuclei \cite{EllV02}, 
as shown in table \ref{ta:T12_2nu}. The most recent entry comes from the Mibeta
group, which built an array of 20 crystals of TeO$_2$\cite{Mibeta03}
of 340 g each, operated as bolometers at a temperature of 8
mK. The energy resolution is excellent, of order $\sigma
(E)/E=1.5\cdot10^{-3}$ at $E_0=2528$~keV. Most crystals use natural tellurium
(34.5 \% $^{130}$Te). But recent measurements were
performed with two of the crystals enriched in $^{130}$Te, and two
more enriched in $^{128}$Te. The comparison yields a positive signal,
corresponding to the half-life reported in table \ref{ta:T12_2nu}
for $^{130}$Te, and 
in reasonable agreement with earlier geochemical results.
 
\begin{table*}[t]
\begin{center}
\begin{tabular}{|r|c|c|c|}
\hline
  & $T_{1/2}^{2\nu} meas.$ & $\frac{T_{1/2}^{calc}} {T_{1/2}^{meas}}$ & 
$\frac{T_{1/2}^{calc}} {T_{1/2}^{meas}} $ \\ 
 & (yr)& QRPA\cite{Eng88,MoeV94} & Shell model \cite{Caur96}\\ \hline
\begin{tabular}{r} 
$^{48}$Ca  \\ 
$^{76}$Ge  \\ 
$^{82}$Se  \\ 
$^{100}$Mo \\ 
$^{130}$Te \\ 
\end{tabular} 
&\begin{tabular}{c} 
$(4.2\pm 1.2)\cdot 10^{19}$\\
$(1.3\pm 0.1)\cdot 10^{21}$\\
$(9.2\pm1)\cdot 10^{19}$\\
$(8\pm0.6)\cdot 10^{18}$\\
$\left(6.1\pm 1.4 stat.\left(^{+2.9}_{-3.5}sys.\right) \right)\cdot 10^{20}$\\
\end{tabular}
&\begin{tabular}{r} 
$-$\\
$1.0$\\
$1.3$\\
$0.76$\\
$0.36$\\
\end{tabular}
 &\begin{tabular}{r} 
$0.91$\\
$2.0$\\
$0.40$\\
$-$\\
$0.38$\\
\end{tabular}
 \\
\hline
\end{tabular}
\caption{List of measured $\beta\beta2\nu$ half-lives; comparison with QRPA and shell model
predictions. \label{ta:T12_2nu}}
\end{center}
\end{table*}
A comparison is made in table \ref{ta:T12_2nu} with calculated
half-lives, using nuclear matrix elements  calculated in the framework
of QRPA \cite{Eng88,MoeV94}, and with the shell model
\cite{Caur96}. Reasonable agreement is observed, at the level of a
factor of 2 or better. 
This is particularly true of the QRPA results,
which shows that the calculations are quite realistic. This encouraged
the authors of ref. \cite{RoF03} to go one step further, and to use the
measured half-lives to fix more precisely the parameters of the QRPA
model,
and to use them to calculate the $\beta\beta0\nu$ matrix elements. These
calculations should be more reliable than older ones, and will be used
in the following to interpret the data on $\beta\beta0\nu$ decay in
terms of $\langle m_{\nu}\rangle$. They tend to give masses slightly
larger than older calculations.

\begin{figure}[htb]
\begin{center}
\hspace*{-0.cm}
\epsfig{file=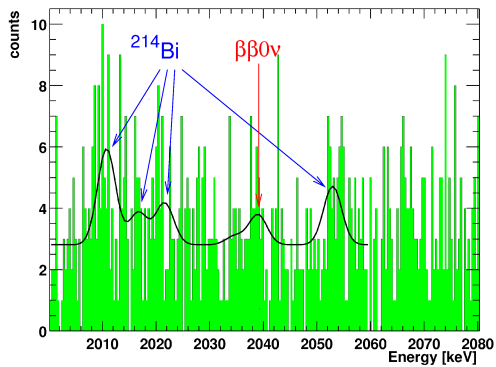,width=7.cm}
\caption{The $\beta\beta0\nu$ energy region in the Heidelberg-Moscow $^{76}$Ge 
data (55 kg$\cdot$yr). The fitted background and $\beta\beta0\nu$ peaks 
are shown.
\label{fi:HMbb0n}}
\end{center}
\end{figure}

Some results on $\beta\beta0\nu$ decay are listed in table
\ref{ta:T12_0nu}, and deduced effective masses $\langle m_{\nu} \rangle$
are given with the QRPA matrix elements of ref. \cite{RoF03} and the shell 
model 
matrix elements of ref. \cite{Caur96}.  They include those of the 
Heidelberg-Moscow
collaboration in Gran Sasso, operating an array of 5 crystals of
germanium (11 kg total mass) enriched to 87 \% in $^{76}$Ge, and
operated as semi-conductor devices, with an energy resolution of order
$\sigma (E)/E=0.7\cdot10^{-3}$ at $E_0=2039$~keV \cite{HM02}. The
advantage of such a good resolution to search for $\beta\beta0\nu$,
and also to identify the background, is illustrated in figure
\ref{fi:HMbb0n}. 
\begin{table*}[tb]
\begin{center}
\begin{tabular}{|r|c|c|c|}
\hline
 & $T_{1/2}^{0\nu}$ [yr] & \multicolumn{2}{c|}{$\langle m_{\nu} \rangle
 $ [eV]}\\    \cline{3-4}
 & (90 \% CL) &  {QRPA}\cite{RoF03} & shell model \cite{Caur96}\\
\hline 
\begin{tabular}{r} 
$^{76}$Ge  \\ 
$^{130}$Te \\ 
$^{136}$Xe 
\end{tabular}
& 
\begin{tabular}{c} 
$(0.8-35)\cdot 10^{25}$ \\ 
$>2.1\cdot10^{23}$ \\ 
$>4.4\cdot 10^{23} $ 
\end{tabular} & 
\begin{tabular}{c} 
$0.13-0.85$ \\ 
$ < 3.6 $ \\ 
$ < 3.3 $ 
\end{tabular} &
\begin{tabular}{c} 
$0.24-1.6$ \\ 
$< -$ \\ 
$< 5.2 $ 
\end{tabular}\\ 
\hline 
\end{tabular}
\caption{Measurements of $\beta\beta 0\nu$ half-lives; Deduced 
effective neutrino 
mass $\langle m_{\nu}\rangle$.} \label{ta:T12_0nu}
\end{center}
\end{table*}
As pointed out by the authors, the background is not flat, as naively
believed previously, also in other experiments, but weak background
peaks from $^{214}$Bi activity ($^{238}$U chain) are visible, in spite
of the high radiopurity of the detector. The peak intensities are
consistent with those of stronger $^{214}$Bi peaks elsewhere in the
spectrum. Fitting the region of interest with these peaks and a
constant background, the authors find indications of an additional
peak centered at $E_0=2039$ eV. They interpret this as an indication of
$\beta\beta0\nu$ decay, corresponding to a half-life of
$T_{1/2}^{0\nu}=(0.8-35)\cdot 10^{25}$~yr (95 \% CL). It would
correspond to a mass $\langle m_{\nu} \rangle$ of order 0.1 to 1.6 eV,
shown in fig. \ref{fi:DBDg}. This result has been the cause of
heated debates, which will not be covered here. What is sure, is that
$\langle m_{\nu} \rangle$ is less than 1.5 eV or so.

Other experiments at present can neither confirm nor 
disprove this indication, as shown in table  \ref{ta:T12_0nu}. Given
are the limits obtained with the already discussed Mibeta bolometers in 
$^{130}$Te, and  in $^{136}$Xe with a xenon gas time projection chamber (TPC) 
operated in the Gotthard underground laboratory \cite{Lues98}. Here
the source consisted in 5.3 kg of xenon enriched to 62.5 \% 
in $^{136}$Xe. The philosophy was somewhat different. In comparison, the
energy resolution is modest ($\sigma (E)/E=2.5\cdot10^{-2}$ at $E_0=2480$
keV). But to a large extent this is compensated by the imaging
capability of the TPC, a powerful tool to identify and suppress
backgrounds, and to select double beta candidates, single continuous tracks
with increased ionization due to larger $dE/dx$ at both ends.
\subsection{Future experiments}
Clearly it is important to have improved data on $\beta\beta0\nu$
decay, first to clear up the indication of the Heidelberg-Moscow
$^{76}$Ge experiment,
and second to extend the sensitivity down to $\langle m_{\nu} \rangle
\simeq 10^{-2}$~eV.  NEMO 3 which has just started taking data in the
Fr\'ejus underground laboratory should be able to perform the first task
\cite{NEMO3}. It consist in a tracking detector with excellent imaging
capability. In contrast to the detectors discussed so far, the source
is not the detector medium. It is a thin foil, placed inside a
fiducial volume filled with helium. This has the advantage that many
nuclei can be investigated, but the disadvantage of energy losses in
the source, deteriorating the  energy
resolution. Nevertheless with a source mass of 7 kg of molybdenum
highly enriched in $^{100}$Mo, and 1 kg of selenium enriched in
$^{82}$Se, it should probe $\langle m_{\nu} \rangle$ down to 0.1 eV or so.

But more ambitious projects are required to go to 
10$^{-2}$ eV. The requirements on future detectors are
much larger masses, of order 1, possibly 10~t, and, at the same time,
much lower background level.

Masses of that kind can be envisioned, since highly efficient
enrichment facilities exist in Russia, based on a variety of
techniques, in particular centrifugation for components which can be
brought to a gaseous form \cite{enrich}.  The reduction in background
level which must go along can be achieved by the use of detector 
components with lower levels of natural and
cosmogenic activities, improved event signature, and finally superior 
energy resolution.

Energy resolution is in any event essential since it is, practically,
the only way allowing to suppress the background from $\beta\beta2\nu$
decay when looking for $\beta\beta0\nu$ decay. The experiments being
considered will search for $\beta\beta0\nu$ half lives of order
$10^{26}$ to $10^{28}$ yr, longer by 5 to 7 orders of magnitude 
comparing with $\beta\beta2\nu$ decay, as depicted in
fig. \ref{fi:bb2n}.  The differences in angular distributions can
help, but not to the same extent. This favors detectors in which the
source is at the same time the detector medium.
\begin{table}[tb]
\begin{center}
\begin{tabular}{|l|l|}
\hline
Majorana & 500 kg $^{76}$Ge, 210 crystals
          in cryostats, segmented readout \\
MOON     & 34 t $^{nat}$Mo sheets between 
          plastic scintillators \\
XMASS    & 10 t of liquid $^{nat}$Xe 
         viewed by photomultipliers \\
CUORE    & 760 kg $^{nat}$TeO$_2$ bolometers \\
GENIUS   & 1(10)t of $^{76}$Ge, in crystals
         suspended in liquid nitrogen \\
EXO      & 1(10)t of $^{136}$Xe, in gas or 
         in liquid TPC, with ion tagging\\
\hline 
\end{tabular}
\caption{Future projects to study $\beta\beta0\nu$ decay, with 
brief description.} \label{ta:futproj}
\end{center}
\end{table}

Several projects are being considered, some as general purpose
detectors for low energy physics, including the study of solar
neutrinos or the search for cold dark matter, others being primarily
dedicated to double beta decay. Table \ref{ta:futproj} gives a non exhaustive 
list. In the following we will briefly discuss three of the projects, which 
will start on a smaller scale, before evolving to the final form.

\underline{CUORE} is a development of Mibeta \cite{Cuore03}. It will
consist in an array of 1000 crystals of $^{nat}$TeO$_2$, with a mass
of 760 g each, operated as bolometers in a single low background
cryostat in the Gran Sasso underground laboratory. 
The energy resolution should be better than 2$\cdot 10^{-3}$
at $E_0=2480$ keV. An elaborate shielding made from roman lead, copper
and PET should protect the crystals against local activities. It is
hoped that the sensitivity to $\beta\beta0\nu$ decay will extend up to
half-lives of order 2$\cdot10^{26}$~yr, corresponding to roughly 0.1
eV in terms of the effective mass $\langle m_{\nu}\rangle$, or somewhat 
better.

In a first step a smaller version, CUORICINO, is being built in the
Gran Sasso laboratory. It will have 56 crystals similar to those
foreseen for CUORE.\\ 
\begin{figure}[htb]
\begin{center}
\hspace*{-0.cm}
\epsfig{file=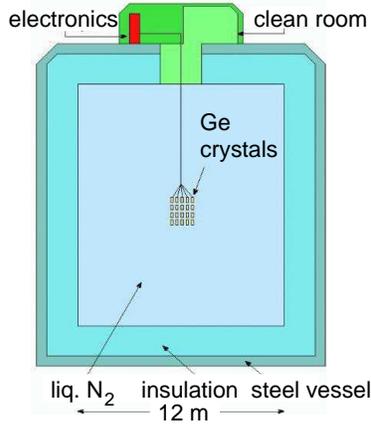,width=5.cm}
\caption{The proposed Genius detector with germanium crystals suspended
in a liquid nitrogen bath.
\label{fi:genius}}
\end{center}
\end{figure}
\underline{GENIUS} has been proposed by the Heidelberg
group \cite{Genius02}. It uses germanium crystals highly enriched in $^{76}$Ge
operated as semiconductor detectors, as in the Heidelberg-Moscow
experiment. But to simplify the design, and reduce the
possibility of radioactive contamination, the crystals are not mounted
in a cryostat, but suspended and immersed in a large vessel filled with liquid
nitrogen, as shown in fig. \ref{fi:genius}. Tests have shown that the
excellent energy resolution inherent to semiconductor detectors can be
maintained in such an arrangement. The FET's of the crystals, a source
of background, can be mounted at a large distance. The liquid nitrogen
can be cleaned to a high level of purity, and provides the innermost
shielding layer against outside activities. Because of the low
density, a large volume is required. Pulse shape discrimination will
be used, as in the later phase of the Heidelberg-Moscow experiment, to
distinguish single site events, potential double beta candidates, from
multiple site events, mostly due to multiple Compton scattering. For
the same reason the crystals will be operated in
anti-coincidence. Careful background estimates have been performed,
taking into account cosmogenic activation, 
leading to encouragingly low rates. The experiment should 
reach a sensitivity of order $10^{-2}$~eV or so to $\langle m_{\nu} \rangle$. A
test facility, with 40 kg of natural germanium, is being set up at Gran
Sasso.\\
\underline{EXO} is devoted to the search of double beta decay in
$^{136}$Xe \cite{EXO00}. To reduce drastically the background the
detector will have a much improved signature. It will be made so as to
observe not only the two electrons emitted in double beta decay, as usual, 
but also to identify the positive ion left behind, in
this case $^{136}$Ba$^{++}$ \cite{Moe91}. It is first brought to the 
singly charged state  $^{136}$Ba$^{+}$. A first 493 nm laser 
brings it from the $6^2S_{1/2}$ ground state to the  $6^2P_{1/2}$ 
excited state. From there it decays back to the ground state, or to
the $5^4D_{3/2}$ metastable state. If this happens, a second 650 nm
laser brings it back to the $6^2P_{1/2}$ state. As long as the
laser irradiation continues, photons are emitted, which allow to
identify the ion. R\&D is in progress to develop this technique.

\begin{figure}[htb]
\begin{center}
\epsfig{file=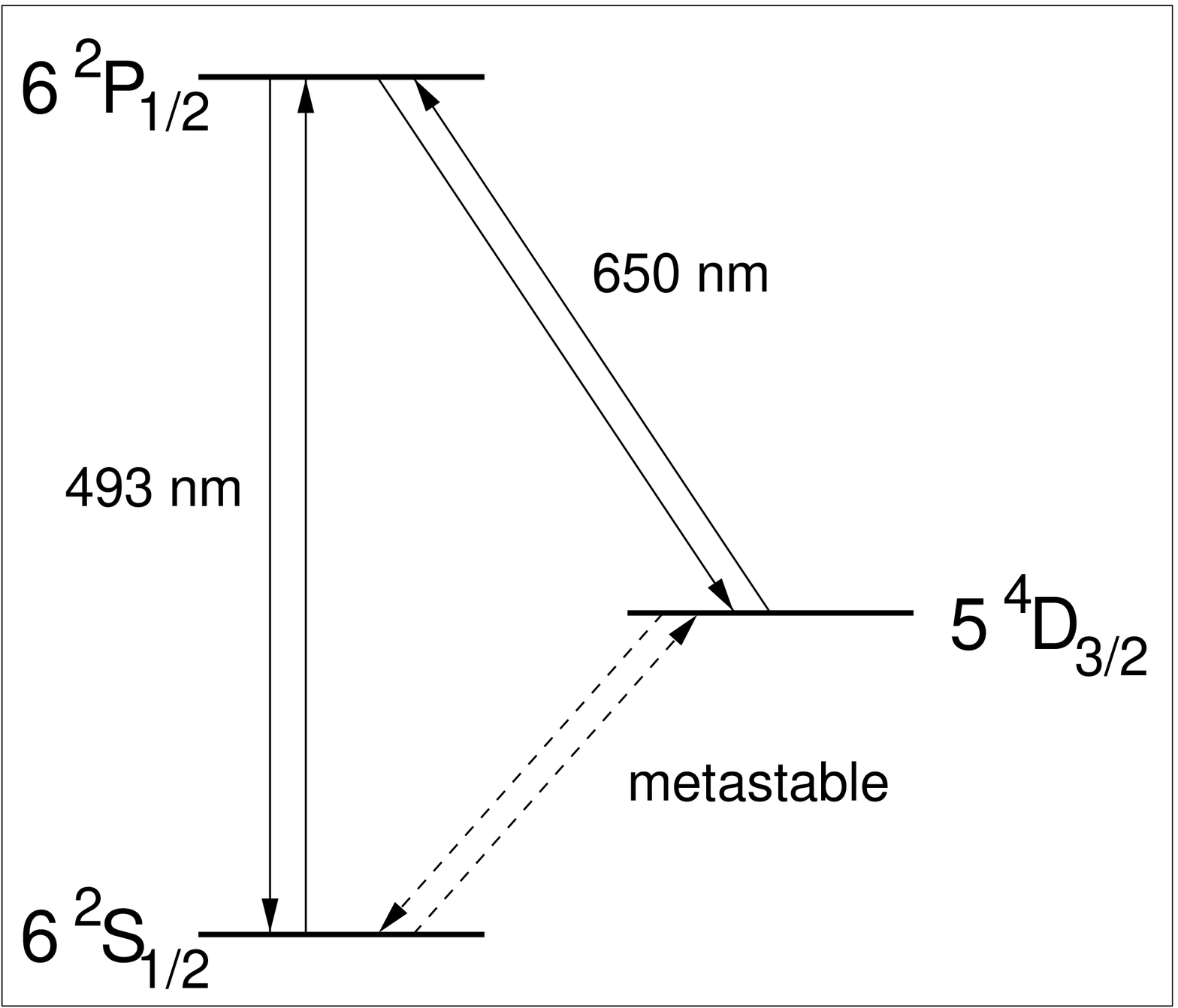,width=5.5cm}
\epsfig{file=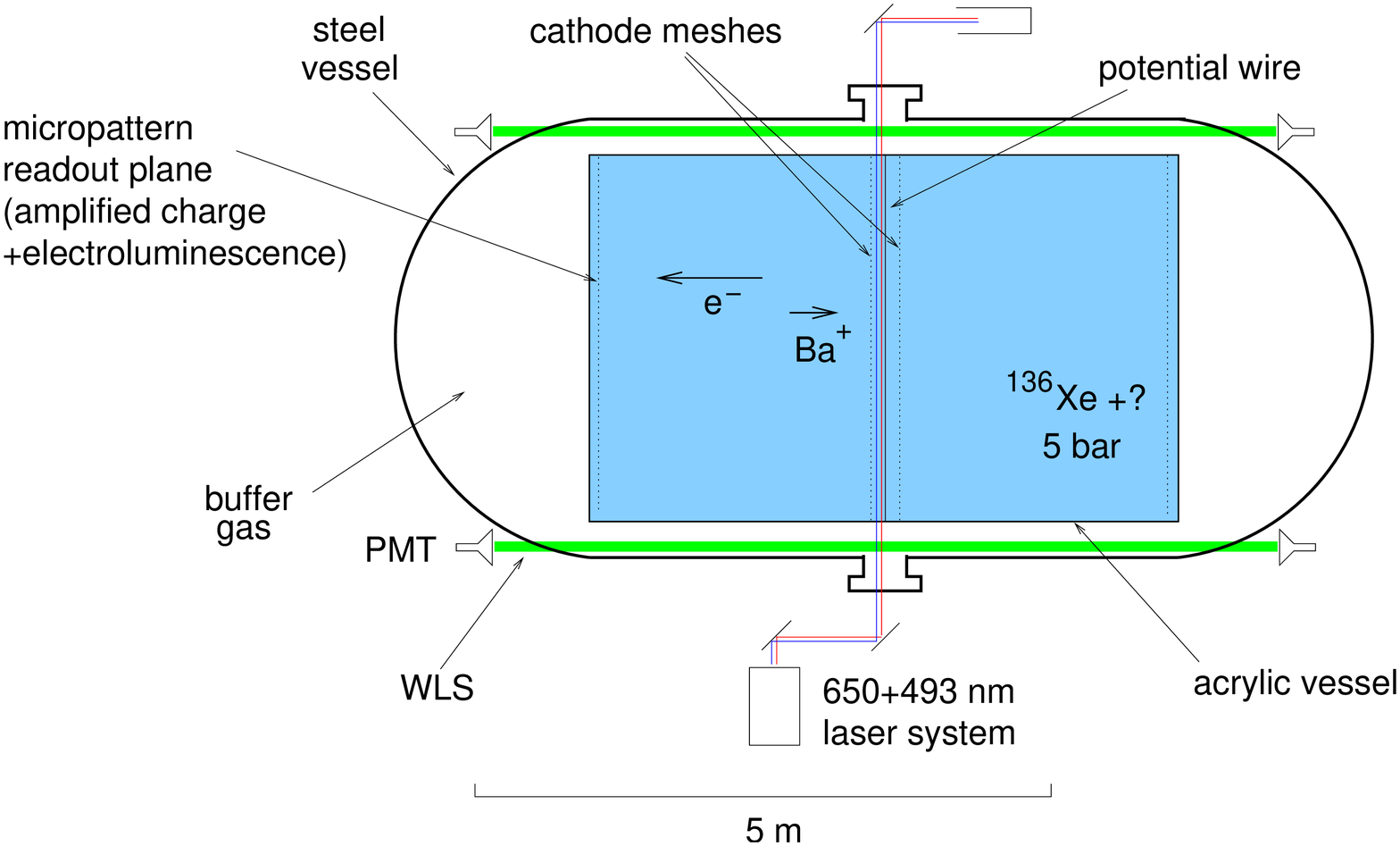,width=8.cm}
\end{center}
\caption{Left: The level scheme of $^{136}$Ba$^+$; right: 
a possible layout for the gas version of the EXO TPC (1~t).
\label{fi:Ba_levels}}
\end{figure}

Two options are being studied for the EXO detector. In either case
the source
consists of xenon highly enriched in $^{136}$Xe and acting as detection 
medium for the electrons. The first possibility is a liquid TPC, which has 
the advantage of being compact and easy to shield. After the detection
of the electrons, by measuring both the scintillation light and the
ionization charge, the much slower positive ion is extracted
with a tip at a negative potential. It is released in an ion trap by
reversing the potential. The ion tagging is then performed in the
trap.

A second possibility is a gas TPC, as in the Gotthard experiment. A
disadvantage is the larger size of the detector. But the advantage is
that the electron tracks are then long enough to be visualized. This
provides a powerful additional selection criterion. The ion tagging
must then be made in the TPC itself, a quencher admixed to the
xenon bringing the ion from the doubly to the singly ionized state. A
possible layout is shown in fig. \ref{fi:Ba_levels}, ressembling that of
MUNU \cite{MUNU}. The TPC is enclosed in 
an acrylic vessel. The ionization electrons drift to the read-out planes, 
at both ends. The positive ion drifts much more slowly toward the cathode 
in the center, near which it crosses the laser beams. The desexcitation 
light is converted to green in wavelength shifter bars, and brought to 
photomultipliers at the ends.

Ion tagging, if it can be made to work efficiently, will lead to an
essentially background free double beta decay experiment. But an
energy resolution of order 2 \% or better is necessary to distinguish
$\beta\beta0\nu$ decay from $\beta\beta2\nu$ decay. This is only
marginally better than what was achieved in the Gotthard experiment
however, and R\&D is in progress for both versions to improve on
that \cite{EXO03}. A first
test experiment will be performed with 50 kg of enriched xenon, but
without tagging. The next goal is an experiment with 1 t, having a 
sensitivity of a few 10$^{-2}$~eV to $\langle m_{\nu} \rangle$. 
A 10 t version could, ultimately, be envisioned.
\section{Conclusion}
Next generation laboratory experiments studying the kinematics in weak decays
and searching for neutrinoless double beta decay will further constrain 
neutrino masses.  They should provide enough information to figure out what
scenario holds for neutrino masses: nearly degenerate, inverted
hierarchy or normal hierarchy.

These experiments are real challenges, facing many
difficulties, in particular fierce background problems. They will thus
proceed in several steps, before reaching the final sensitivity.

\end{document}